\documentclass[a4paper,12pt]{article}
\usepackage[margin=1in]{geometry}

\usepackage{hyperref}
\hypersetup{
	colorlinks=true,
	linkcolor=blue,
	filecolor=magenta,
	urlcolor=cyan,
}

\usepackage{graphicx}
\usepackage{subcaption}
\usepackage{pxfonts}
\usepackage{lineno}

\usepackage{amssymb}
\usepackage{footnote}
\usepackage{multirow}
\usepackage{varwidth}
\usepackage{array}
\usepackage{epstopdf}

\usepackage{color} 

\usepackage{grffile} 
\usepackage{makecell} 
\usepackage[backend=bibtex,url=true,sorting=none,doi=true,maxnames=3,firstinits=true,citestyle=numeric-comp]{biblatex}

\bibliography{library}

\usepackage{lscape}

\usepackage{authblk}

\usepackage{caption}
\DeclareCaptionFormat{myformat}{#1 (color online)#2#3}
\newenvironment{colorfigure}%
{\captionsetup{format=myformat,labelsep=colon}
	\figure
}%
{\endfigure}%

\renewbibmacro*{doi+eprint+url}{%
	\iftoggle{bbx:doi}
	{\printfield{doi}}
	{}
	\newunit\newblock
	\iftoggle{bbx:url}
	{\iffieldundef{doi}{
			\iffieldundef{arxivId}{	\printfield{url}}{}}{}
	}
	{}%
	\newunit\newblock
	\iftoggle{bbx:eprint}
	{\usebibmacro{eprint}}
	{}%
}

\title{Opportunities for Two-color Experiments at the SASE3 undulator line of the European XFEL}

\author[1]{Gianluca Geloni}
\author[2]{Vitali Kocharyan}
\author[1]{Tommaso Mazza}
\author[1]{Michael Meyer}
\author[2]{Evgeni Saldin}
\author[1]{Svitozar Serkez}
\affil[1]{European XFEL GmbH, Hamburg, Germany}
\affil[2]{Deutsches Elektronen-Synchrotron (DESY), Hamburg, Germany}

\date{}

\begin{document}
\thispagestyle{empty}
\begin{Large}
\textbf{DEUTSCHES ELEKTRONEN-SYNCHROTRON}

\textbf{\large{Ein Forschungszentrum der
Helmholtz-Gemeinschaft}\\}
\end{Large}

DESY 17-068

May 2017

\begin{eqnarray}
\nonumber &&\cr \nonumber && \cr \nonumber &&\cr
\end{eqnarray}
\begin{eqnarray}
\nonumber
\end{eqnarray}
\begin{center}
\begin{Large}
\textbf{Opportunities for Two-color Experiments at the SASE3 undulator line of the European XFEL}
\end{Large}
\begin{eqnarray}
\nonumber &&\cr \nonumber && \cr
\end{eqnarray}

\begin{large}
Gianluca Geloni, Tommaso Mazza, Michael Meyer and Svitozar Serkez
\end{large}
\textsl{\\European XFEL GmbH, Hamburg}
\begin{large}

Vitali Kocharyan and Evgeni Saldin
\end{large}
\textsl{\\Deutsches Elektronen-Synchrotron DESY, Hamburg}
\begin{eqnarray}
\nonumber
\end{eqnarray}
\begin{eqnarray}
\nonumber
\end{eqnarray}
ISSN 0418-9833
\begin{eqnarray}
\nonumber
\end{eqnarray}
\begin{large}
\textbf{NOTKESTRASSE 85 - 22607 HAMBURG}
\end{large}
\end{center}
\clearpage
\newpage

\maketitle
	

\begin{abstract}
X-ray Free Electron Lasers (XFELs) have been proven to generate short and powerful radiation pulses allowing for a wide class of novel experiments. If an XFEL facility supports the generation of two X-ray pulses with different wavelengths and controllable delay, the range of possible experiments is broadened even further to include X-ray-pump/X-ray-probe applications. In this work we discuss the possibility of applying a simple and cost-effective method for producing two-color pulses at the SASE3 soft X-ray beamline of the European XFEL. The technique is based on the installation of a magnetic chicane in the baseline undulator and can be accomplished in several steps. We discuss the scientific interest of this upgrade for the Small Quantum Systems (SQS) instrument, in connection with the high-repetition rate of the European XFEL, and we provide start-to-end simulations up to the radiation focus on the sample, proving the feasibility of our concept.

\end{abstract}

%
%

%


\section{ Method }

\begin{figure}
	\centering
	\begin{subfigure}[t]{1.0\textwidth}
		\centering
		\includegraphics[width=1.0\textwidth]{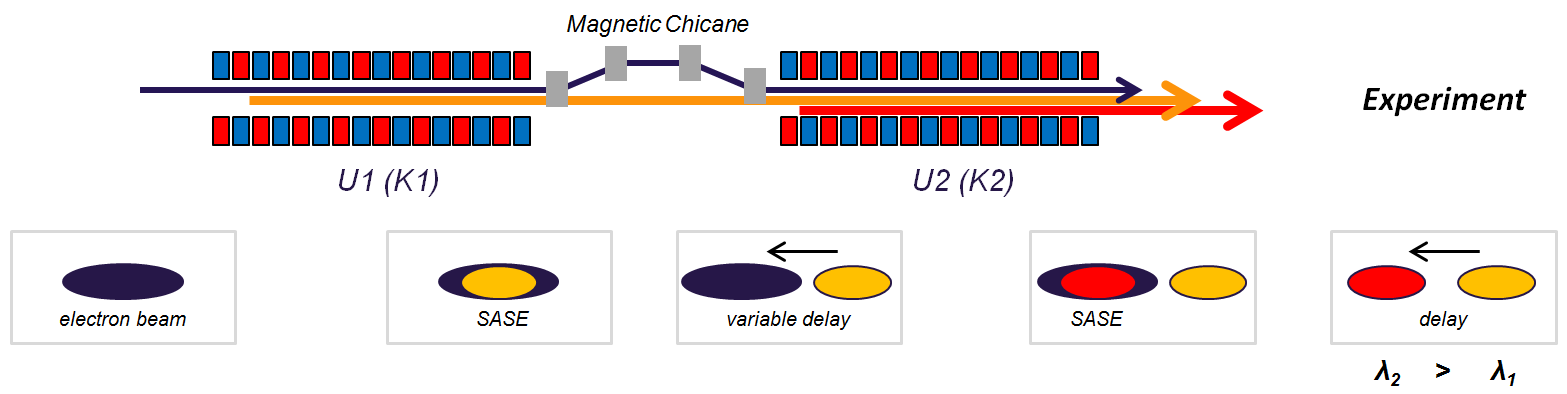}
		\caption{Chicane only}
	\end{subfigure}
	\\ \vspace{0.5in}
	
	\begin{subfigure}[t]{1.0\textwidth}
		\centering
		\includegraphics[width=1.0\textwidth]{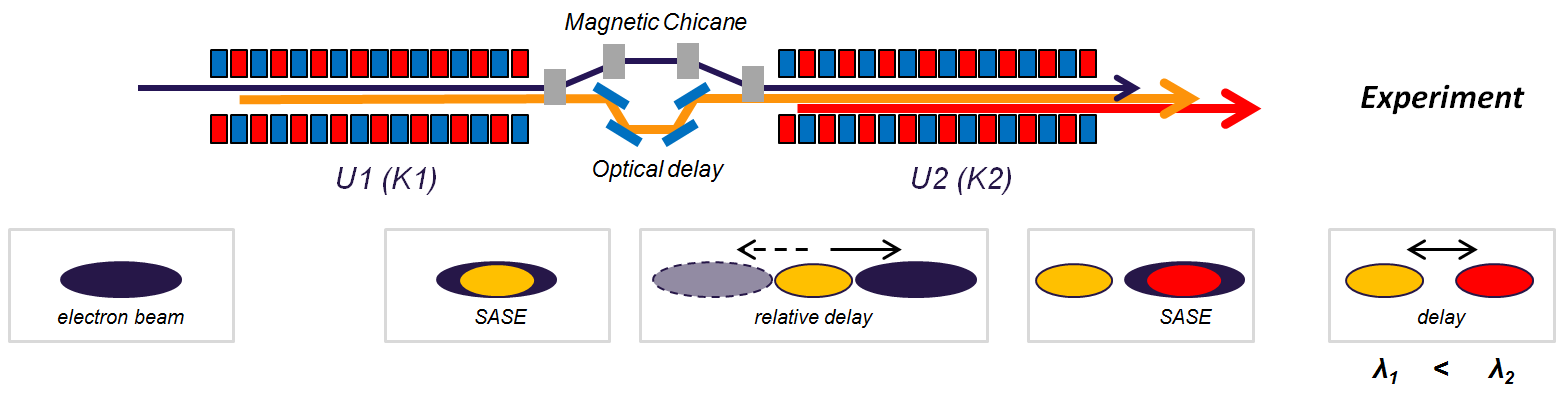}
		\caption{Chicane and optical delay line}
	\end{subfigure}
	\caption{(color online) Schematic illustration of a simple two-color FEL technique without (top) and with the addition of a compact optical delay line (bottom).}
	\label{scheme1}
\end{figure}

The simplest way currently available to enable the generation of two closely separated (in the order of 50\,fs) pulses of different wavelengths (later - colors) at X-ray Free-Electron lasers consists of inserting a magnetic chicane between two undulator parts as suggested in~\cite{Geloni2010b} and experimentally proven in~\cite{Lutman2013,Hara2013}. The scheme is illustrated in Figure~\ref{scheme1}-a. We propose to split the baseline SASE3 soft X-ray undulator into two parts with a magnetic chicane. Both parts act as independent undulators and will be referred further as $ U1 $ and $ U2 $. The nominal electron beam enters the first undulator $U1$,  tuned to the resonant wavelength $\lambda_1$. After passing through $ U1 $, both electron beam and emitted radiation enter the chicane. This magnetic chicane has two functions: first, it introduces a suitable delay between the electron beam and the radiation generated in $U1$. Delays from zero\footnote{In our case, due to radiation slippage in the subsequent undulators, the effective minimum delay between the two pulses of different colors is of the order of several femtoseconds.} up to the picosecond level can be obtained with a compact magnetic chicane of several meters length. Second, due to dispersion, the passage of the electron beam through the magnetic chicane smears out the microbunching at wavelength $\lambda_1$. As a result, when the -after the magnetic chicane- delayed electron beam enters the second undulator $U2$, the Self-Amplified Spontaneous Emission (SASE) process starts from shot-noise again. Therefore, if the undulator $U2$ is tuned to the resonant wavelength $\lambda_2$, then at the undulator exit one obtains a first radiation pulse at wavelength $\lambda_1$ followed by a second one with wavelength $\lambda_2$ delayed by a time interval that can be varied by changing the strength of the chicane magnets.

One must ensure that the electron beam quality at the entrance of the second undulator $ U2 $ is still good enough to sustain the FEL process. This poses limits on the maximum power that can be extracted from $U1$. In particular, the amplification process there should not reach saturation. Optimization of the maximum power also poses limits on the wavelengths choices The wavelength separation between the two pulses can theoretically span across the entire range made available by the undulator system, in the case of SASE3 between about $250$ eV and $3000$ eV. However, the impact of the FEL process on the electron beam quality depends on the radiation wavelength. Therefore, in order to maximize the combined radiation power that can be extracted, especially at large wavelength separations, the first pulse to be produced should be at the shortest wavelength. Moreover, the magnetic chicane strength should be large enough to smear out the microbunching at $\lambda_1$, unless the separation between $\lambda_1$ and $\lambda_2$ is larger than the FEL bandwidth.

An easy way to increase the flexibility of the scheme is to introduce a compact optical delay line to have full control on the relative temporal separation between the two pulses as shown in Figure~\ref{scheme1}-b. Since the photon beam transverse size at the position of the magnetic chicane is, roughly speaking, as small as the electron beam, i.e. a few tens of microns, the length of each mirror can be as short as several centimeters. In order to simplify the design of the mirror delay line, one may fix the optical delay to a few hundred femtoseconds, thus avoiding the use of moving mirrors, and subsequently tune the delay by changing the current in the magnetic chicane coils. Therefore, the introduction of an optical delay line would allow one to sweep between negative and positive delays at a cost of a smaller delay tuneability, caused by a lower limit of the chicane magnetic field (while the optical delay is inserted).

It is also worth noting that the scheme proposed here can, in principle, be upgraded to a self-seeding setup in the soft X-ray range (SXRSS), by inserting a SXRSS monochromator in place of the optical delay line, and by properly choosing the lengths the $U1$ and $U2$ undulators. When the SXRSS setup is retracted, one still has the possibility of producing 2-color pulses as in Figure~\ref{scheme1}. Such flexible use of a magnetic chicane is not new. In fact, the two-color mode of operation at the LCLS~\cite{Lutman2013} and SACLA (for hard X-rays)~\cite{Hara2013} makes use of magnetic chicanes that are parts of self-seeding setups. In this way, however, the introduction of an optical delay as in Figure~\ref{scheme1}-b becomes more complicated. One needs, in fact, to consider modifications to the SXRSS monochromator allowing for the introduction of the optical delay already in the SXRSS hardware, that would then work both as monochromator or, when needed, as optical delay.

Finally, one can think of equipping the SASE3 undulator with a second magnetic chicane to enable, for instance,  fresh-bunch techniques or a combination of seeded and SASE pulses with different colors.

Even the simplest way of generating two-color pulses at the SASE3 beamline of the European XFEL, in combination with the high-repetition rate capabilities of the facility is expected to enable novel exciting science at the two soft X-ray instruments: Small Quantum Systems (SQS)~\cite{Mazza2012} and Spectroscopy \& Coherent Scattering (SCS)~\cite{Scherz2013}.

In the following sections we limit ourselves to the analysis of one science case for the SQS  instrument, including FEL simulations and wavefront propagation studies up to the sample, while leaving further studies for the SQS and the SCS instruments for the future.

\section{SQS science case}\label{sec:sqs_sci_case}

The two-color operation mode enables a large number of scientific applications based on a pump-probe excitation scheme with two individually controllable X-ray pulses. In the following a concrete example is discussed, which will make use of site-specific excitation in molecules possible at SQS scientific  instrument of the European XFEL~\cite{Mazza2012}. The excitation of a specific atomic site is enabled using soft X-ray pulses, since the radiation efficiently couples to the strongly bound core electrons, which are localized at the atomic site. Tuning the wavelength of the pump pulse to a specific threshold, a molecule can be excited at a well-defined atomic position. Using then the probe at another wavelength, which is connected to another core hole excitation, possible changes induced by the first pulse at a different site in the molecule are measured. Finally, the variation of the time delay between the two pulses provide access to the dynamics of this process, i.e. on the time, which is necessary to transport the information from one position in the molecule to another.

As illustrating example, we discuss charge transfer processes in a linear molecule, such as $I-C_n-H_{2n}-Cl$, composed of long carbon chain with two different halogen atoms, e.g. iodine and chlorine, at both ends (Figure~\ref{fig:charge_transfer}). In the wavelength range accessible with the SASE3 undulator the $2p$ core electron of chlorine (threshold at 210\,eV) as well as the $3d$ core electron of iodine (threshold at 630\,eV) can be ionized. Considering a first pulse (pump) at a photon energy of 250\,eV, the perturbation introduced by the XUV photon will be localized at the chlorine site, since core electrons of the other atoms ($I$ and $C$) are still not in reach and cross section for valence ionization is weak. In the same way, choosing for the probe pulse the second photon energy at 630\,eV assures that preferentially (i.e. most efficiently) the $3d$ electron at the iodine site is excited or ionized.

\begin{colorfigure}
	\centering
	\includegraphics[width=0.5\textwidth]{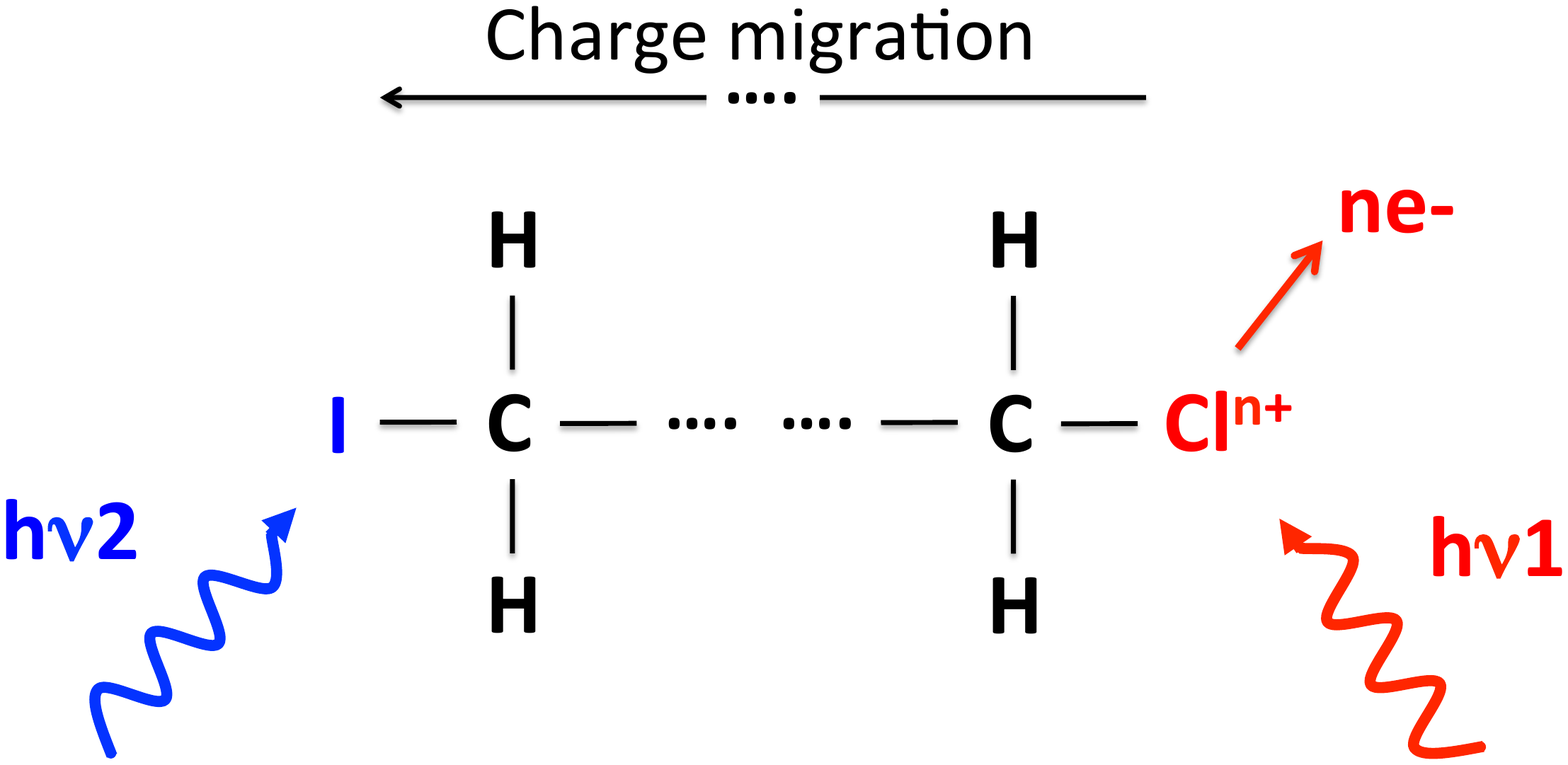}
	\caption{Schematic representation of the two-color pump-probe process in molecular $I-C_n-H_{2n}-Cl$.}
	\label{fig:charge_transfer}
\end{colorfigure}

An efficient and informative experimental method to monitor the intramolecular processes is given by high-resolution Auger spectroscopy, ideally in combination with ion spectroscopy performed in a coincidence arrangement.  The $3d$ Auger spectrum of iodine is located in the kinetic energy range around 400-500\,eV arising mainly from the most prominent transitions to doubly charged states with electron configurations $4d^{-2}$ and $4d^{-1}4p^{-1}$~\cite{Jonauskas2003}. These lines are well separated from the corresponding $Cl~2p$ Auger spectrum at kinetic energies between 165 and 175\,eV~\cite{Kivilompolo2000} and other ionization processes taking place at the photon energies considered here. An illustration of all possible ionization processes, including also the valence ionization of all atoms at both wavelengths is given in Figure~\ref{fig:electron_spectrum}.

\begin{colorfigure}
	\centering
	\includegraphics[width=0.9\textwidth]{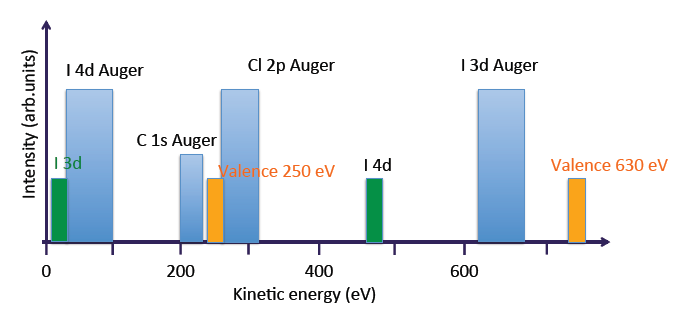}
	\caption{Schematic representation of the electron spectrum recorded upon ionization with two pulses at photon energies at 250 and 630~eV, respectively}
	\label{fig:electron_spectrum}
\end{colorfigure}

Due to the high intensity of the FEL pulses sequential ionization processes are possible and likely to happen. As a consequence, the electron spectrum of the neutral parent molecule (as depicted in Figure~\ref{fig:electron_spectrum}) will be overlaid with emission lines arising from the ionization of the ionic species and of the dissociation fragments. In order to separate the emission from different species coincidences between electrons and ionic fragment can be used for a more detailed analysis. In fact, coincidence experiments will be one of the major experimental tools available at the SQS instrument, and are feasible due to the high number of X-ray pulses (up to 27 000 per second) at the European XFEL. This high repetition rate allows one to record data of high statistics for coincident measurements between electrons and ionic fragments coming unambiguously from the same molecule.

In a typical experimental scenario, first the Auger spectra would be recorded at the individual wavelength 250\,eV and 630\,eV, respectively, in order to obtain the one-photon reference spectra. In addition, electron-ion coincidence will provide charge and fragment resolved electron spectra would at these photon energies. In a second step, the Iodine $3d$ Auger spectrum -- caused by the 630\,eV photon pulse -- will be monitored in the presence of the additional the 250\,eV pulse. When the 250\,eV pulse comes after the 630\,eV pulse, the spectrum will be unchanged compared to the single color spectrum. When both pulses are overlapping or the 250\,eV comes earlier, the observation of the iodine Auger spectrum for different delays between both pulses provide the information about the intermolecular processes. Changes of the kinetic energy position and of the intensity distribution within the $I 3d$ Auger spectrum are the monitor to follow charge migration processes inside the molecule, i.e. to determine e.g. the time required to transmit the information about the creation of a $2p$ core hole on the chlorine site to the iodine atom. For small molecules this time scale is in the order to a few femtoseconds~\cite{Golubev2015}, so probably difficult to access with pulses of about 2\,fs duration each. For longer carbon chains the time scale is expected to increase to about 10\,fs or more and therefore well suited to be studied with the set-up at the SQS instrument.

Furthermore, by selecting in coincidence mode a fragment containing the iodine atom or the iodine atom itself, also information on charge transfer processes is made available. Compared to earlier work using an optical laser to initiate the fragmentation~\cite{Erk2014}, the pump can be used in a very selective way, changing for example between excitations of the chlorine and the carbon atom. In this way a more versatile and detailed analysis of the complex intra-molecular interaction and on the related electron and nuclear dynamics will become possible.

\section{Simulations for the SQS science case}

For the particular science case at the SQS instrument discussed in the previous section, two fs-order-long X-ray pulses with a tunable relative delay are required. In what follows we consider a simulation scenario where a magnetic chicane and an optical delay line are installed at SASE3, see Figure~\ref{fig:scheme}.

\begin{colorfigure}
	\centering
	\includegraphics[width=1.0\textwidth]{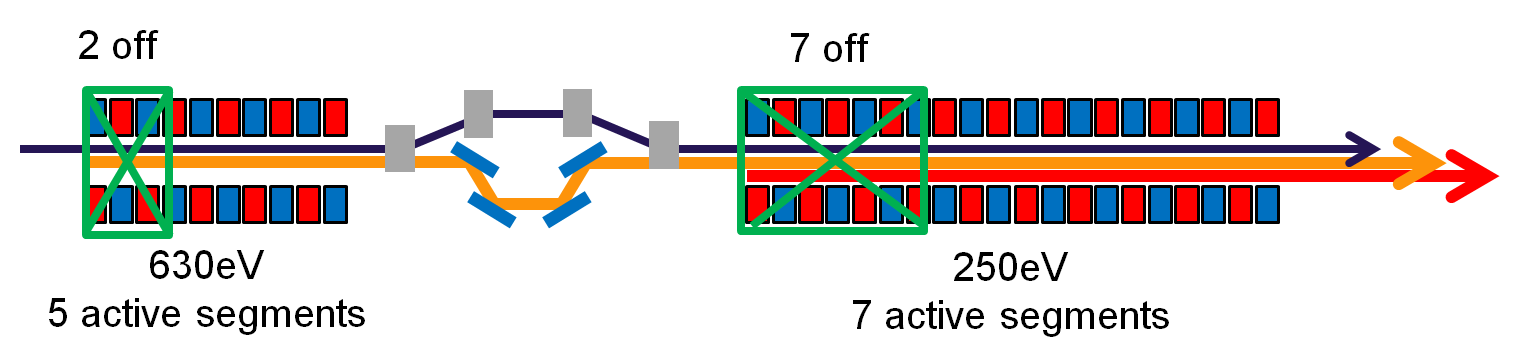}
	\caption{The SASE3 undulator, consisting of 21 undulator segments. It is effectively separated into two parts U1 and U2 by the magnetic chicane with an optical delay line.}
	\label{fig:scheme}
\end{colorfigure}

\begin{colorfigure}
	\centering
	\includegraphics[width=0.8\textwidth]{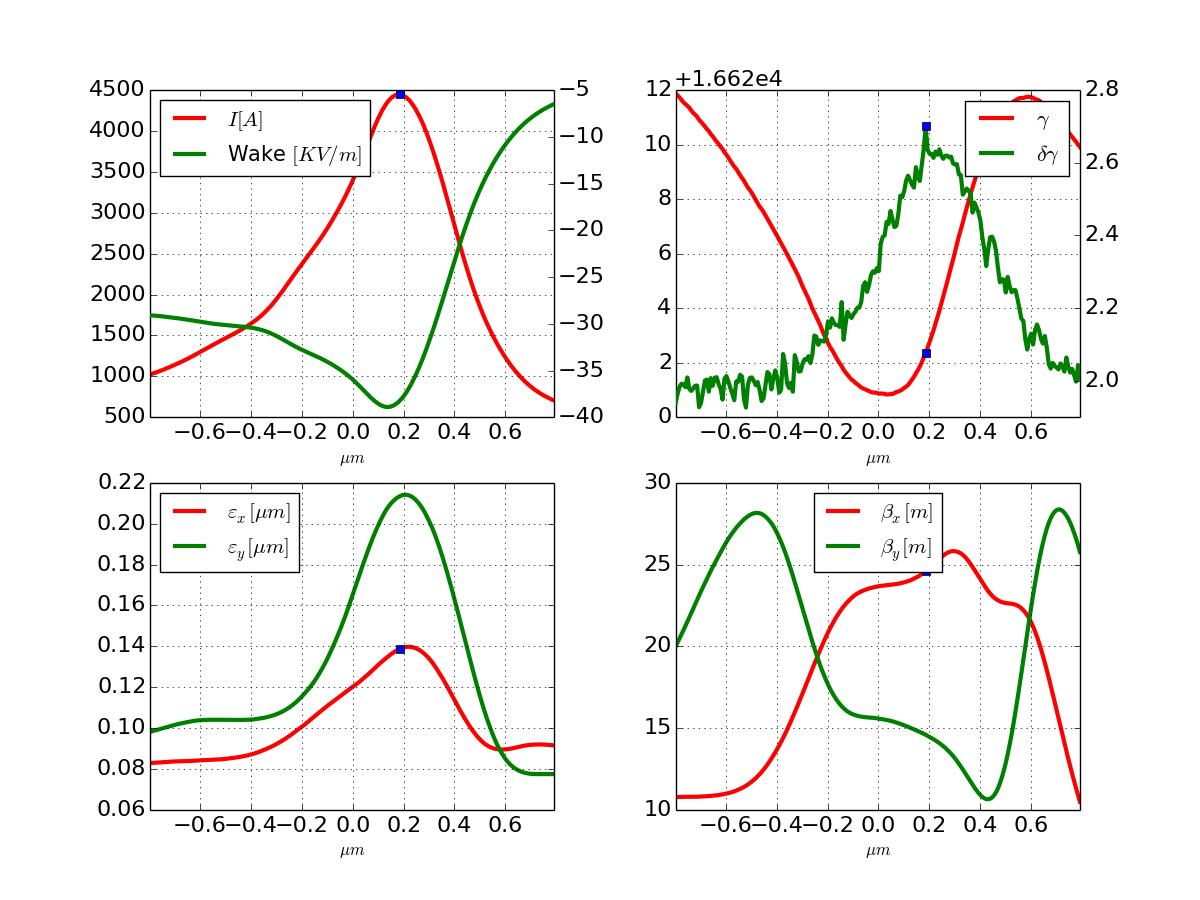}
	\caption{Nominal 20pC electron beam at the entrance to the SASE3 undulator}
	\label{fig:ebeam}
\end{colorfigure}
Figure~\ref{fig:ebeam} shows the result of start-to-end simulations for the electron beam through the European XFEL linac to the entrance of the SASE3 beamline based on \cite{mpy_web}. In order to deliver the electron beam to the SASE3 undulator line, it should pass through the SASE1 line. It was shown~\cite{Brinkmann2010}, that lasing in the SASE1 undulator can be inhibited, but due to quantum fluctuations~\cite{Rossbach1996,Rossbach1997} and synchrotron radiation the electron parameters deteriorate. Also, the influence of the resistive wake in the SASE1 undulator affects the electron beam energy distribution. Therefore, we simulated the electron beam propagation through the SASE1 undulator, accounting for all these effects.

\begin{figure}[!htb]
	\centering
	\includegraphics[width=0.5\textwidth]{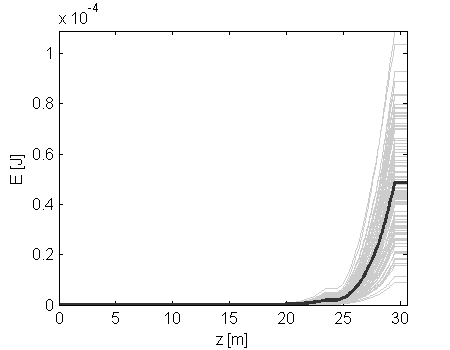}
    \includegraphics[width=0.5\textwidth]{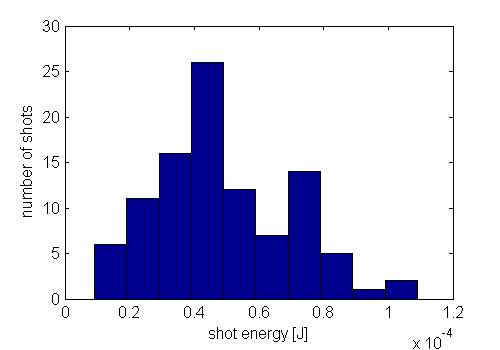}
    \includegraphics[width=0.5\textwidth]{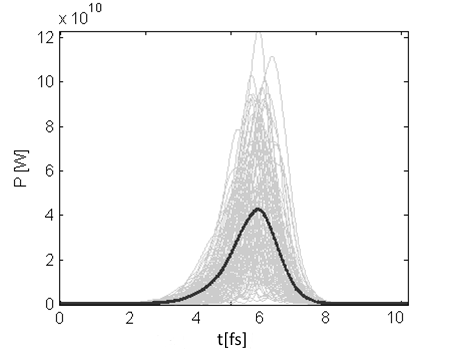}
    \includegraphics[width=0.5\textwidth]{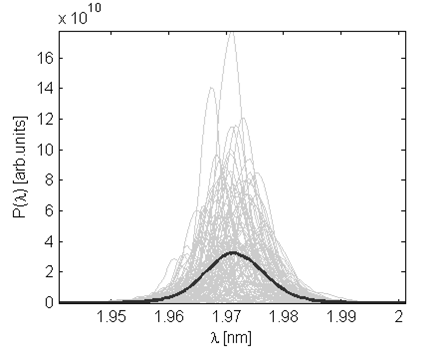}
	\caption{The radiation properties from $ U1 $ tuned at the fundamental photon energy of 630~eV: pulse energy growth along the undulator (top left), distribution of FEL shots with different energies per pulse (top right), power along the FEL pulses (bottom left) and their spectra (bottom right). Mean energy per pulse is $50~\mu$J, corresponding to $5\times10^{11}$ photons per pulse on average.}
	\label{fig:SASE_630}
\end{figure}

The electron beam obtained in this way is sent through the first part of the SASE3 undulator $ U1 $ (see Figure~\ref{fig:scheme}), which is composed of 7 segments, with the first two segments not contributing to the SASE process and the rest 5 segments tuned at 630~eV. In order to inhibit SASE emission we can simply open the undulator gap until no lasing is sustained. The results of a statistical simulation run calculated by Genesis~\cite{Reiche2004}, which has been used for all simulations in this paper, are shown in Figure~\ref{fig:SASE_630}.

\begin{figure}[!htb]
	\centering
	\includegraphics[width=0.5\textwidth]{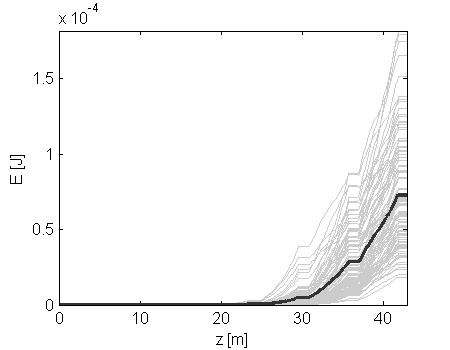}
    \includegraphics[width=0.5\textwidth]{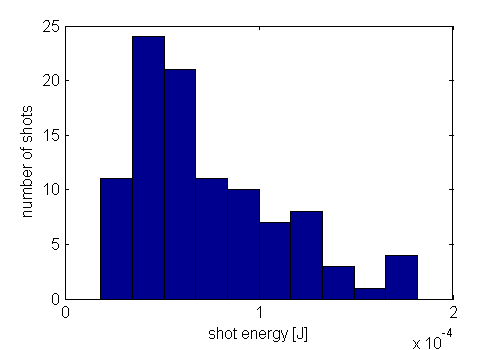}
    \includegraphics[width=0.5\textwidth]{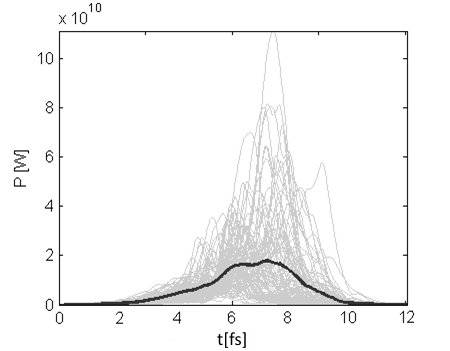}
    \includegraphics[width=0.5\textwidth]{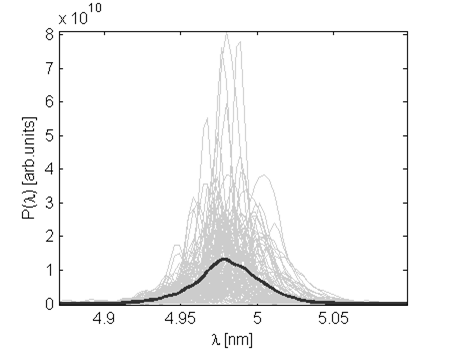}
	\caption{The radiation properties from $ U2 $ tuned at the fundamental photon energy of 250~eV: pulse energy growth along the undulator (top left plot), distribution of FEL shots with different energies per pulse (top right), power along the FEL pulses (bottom left) and their spectra (bottom right plot).  Mean energy per pulse is $70~\mu$J, corresponding to $1.7\times10^{12}$ photons per pulse on average.}
	\label{fig:SASE_250}
\end{figure}
The photon beam then passes through the fixed optical delay line, while the electron beam goes through the magnetic chicane. The delay in the magnetic chicane can be adjusted with sub-fs accuracy, and can be set to under- or over-compensate the optical delay. In this way any delay between the two photon pulses, positive or negative, can be set. If the desired color separation is smaller than 1\%, i.e. when both colors are within the SASE amplification bandwidth, we should set the magnetic chicane to have a dispersion strength large enough to destroy the microbunching developed in the first undulator part. We work under the assumption that X-ray diffraction effects in the optical delay line are negligible and that its mirrors do not sensibly modify the radiation wavefront distribution. 
Moreover, we estimate that a variable delay up to several hundreds femtoseconds with the before-mentioned accuracy of a fraction of femtosecond can be provided between the two radiation pulses. Due to the introduction of the optical delay line into the magnetic chicane, the 630\,eV radiation pulse generated in the $ U1 $ undulator is delayed with respect to the electron beam at the $ U2 $ undulator entrance. The electron beam propagated through the chicane in our simulations maintains all its parameters, except the electron density modulation (microbunching). We dump the numerically simulated electron beam distribution at the end of undulator $ U1 $ and use it for the FEL simulations in $ U2 $. The proper shot noise is automatically introduced into the electron beam by the Genesis simulation code. The second undulator $ U2 $ consists of 14 segments, with the first seven switched off and the remaining seven lasing at 250\,eV. The results of a $ U2 $ statistical simulation run are presented in Figure~\ref{fig:SASE_250}.

\begin{colorfigure}[!htb]
	\centering
	\includegraphics[width=0.7\textwidth]{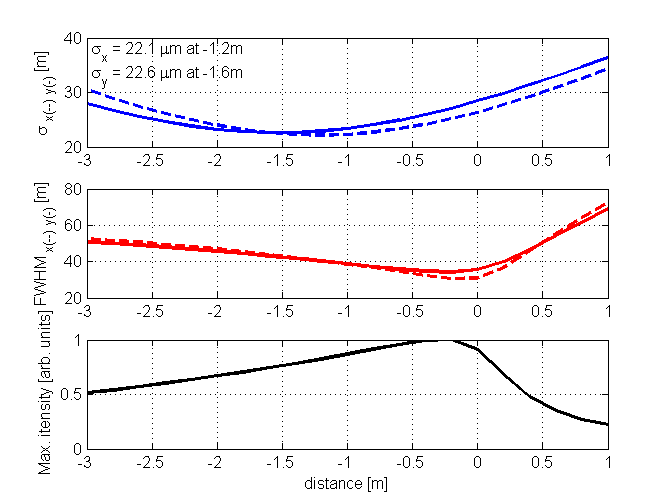}
	\caption{Source analysis by backpropagation at 630 eV.}
	\label{fig:backprop_630}
\end{colorfigure}

It is important to analyze the properties of the FEL radiation in order to be able to simulate its propagation through the optical beamline. In fact, the longitudinal position of the source jitters from shot to shot due to the statistical properties of SASE. An average position is found by choosing the radiation shot most similar to the average one, and back-propagating  the radiation field in free-space. The back-propagation is performed with an in-house radiation propagation code. Results are presented in Figure~\ref{fig:backprop_630} for the $630$~eV radiation pulse case. Similar results hold for the pulse at $250$~eV. As it can be seen, there remains some freedom in selecting the actual position of the source. In fact, since the source is non-Gaussian, the position where the photon density reaches its maximum does not correspond to the position of the minimum FWHM or RMS spot size. It should be noted however, that due to the large magnification factor at the sample position, variations within a few meters are of no significant importance. Here we define the position of the source as the point of the maximum photon density.

%
The final step consists in performing wavefront propagation simulations thorough the SQS beamline. A full pulse propagation is possible, where a simulated FEL pulse is propagated, but this procedure does not add particular insight to the final results. Therefore, in order to simplify the propagation analysis, the source is modeled with a Gaussian beam of the same divergence.

\begin{colorfigure}[!htb]
	\centering
	\includegraphics[width=1.0\textwidth]{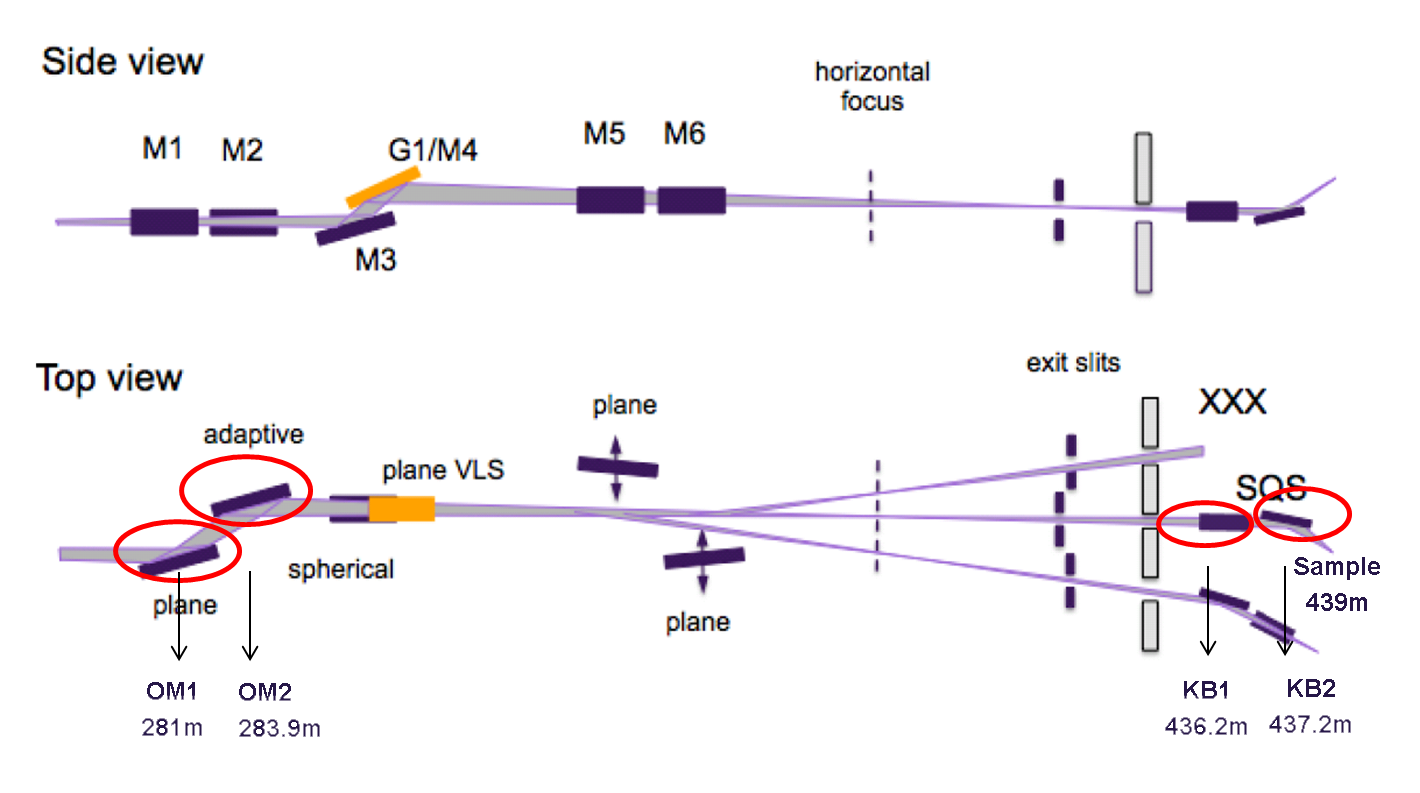}
	\caption{Optical elements relevant to SQS and their positions assumed for calculations}
	\label{fig:SQS_line}
\end{colorfigure}
%

\begin{colorfigure}[!htb]
	\centering	
	\includegraphics[width=0.7\textwidth]{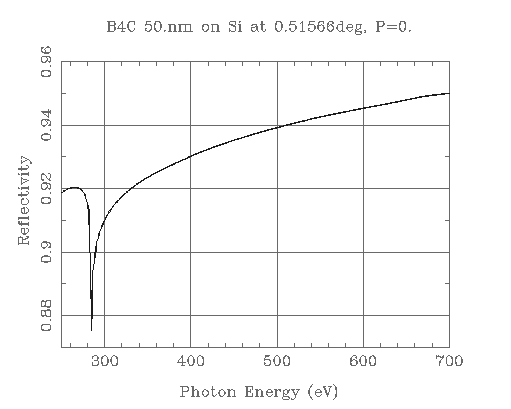}
	\caption{Reflectivity of B4C on Si substrate at $9$ mrad, as a function of the photon energy ~\cite{CXRO}.}
	\label{fig:reflectivity}
\end{colorfigure}

We model the SQS beamline with two offset mirrors and two KB mirrors, Figure~\ref{fig:SQS_line}. For simplicity, in this example we chose the same incident angle of 9\,mrad for all mirrors, though 12\,mrad is possible for the offset mirrors. The reflectivity is then fixed, see Figure~\ref{fig:reflectivity}. We assume 1\,nm RMS height errors for the mirror surfaces.


\begin{colorfigure}[!htb]
	\centering
	\includegraphics[width=0.75\textwidth]{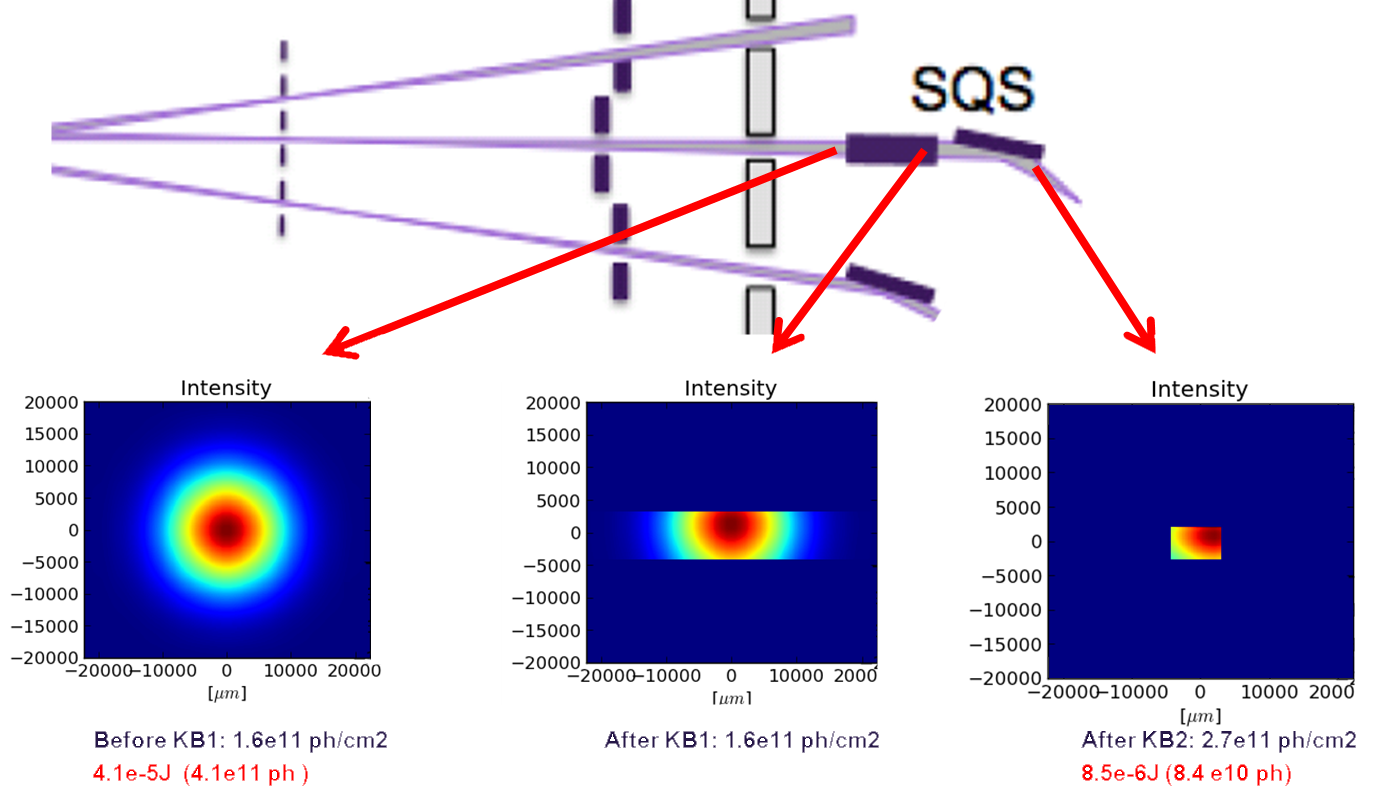}
	\caption{Illustration of radiation clipping due to the limited length of the refocusing KB mirrors. The example is based on 630eV photon energy. Nearly 80\% of the radiation does not pass the KB mirror pair.}
	\label{fig:clip}
\end{colorfigure}

The finite length of the mirrors, together with the grazing incidence angle and the long propagation distance are potentially responsible for clipping of the radiation pulse along the transverse directions and, consequently, for diffraction effects. In fact we treat the finite length of mirrors as effective apertures through which radiation is propagated. We did not observe any visible effects originating from the offset mirrors, while the KB mirror pair introduces a significant clipping, illustrated on Figure~\ref{fig:clip}. It should be remarked here that the use of the intermediate focus that can be generated by the second offset mirror would reduce the beam footprint on the second KB and thereby increase the geometric transmission. The performances of this option are not calculated in the present work.

\begin{colorfigure}[!htb]
	\centering
	\includegraphics[width=0.75\textwidth]{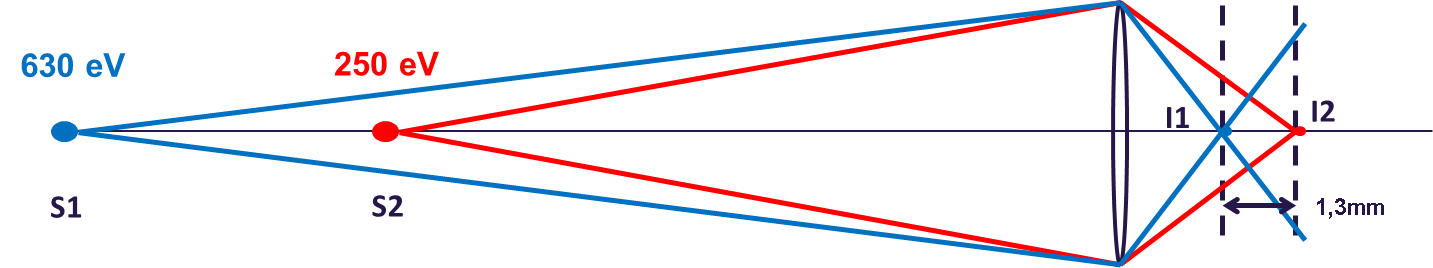}
	\caption{Illustration of the two spatially separated sources $ S1 $ and $ S2 $ being reimaged with a single focusing element (for simplicity). Each source is reimaged at a different position.}
	\label{fig:2sources}
\end{colorfigure}

\begin{colorfigure}[!htb]
	\centering
	\includegraphics[width=0.7\textwidth]{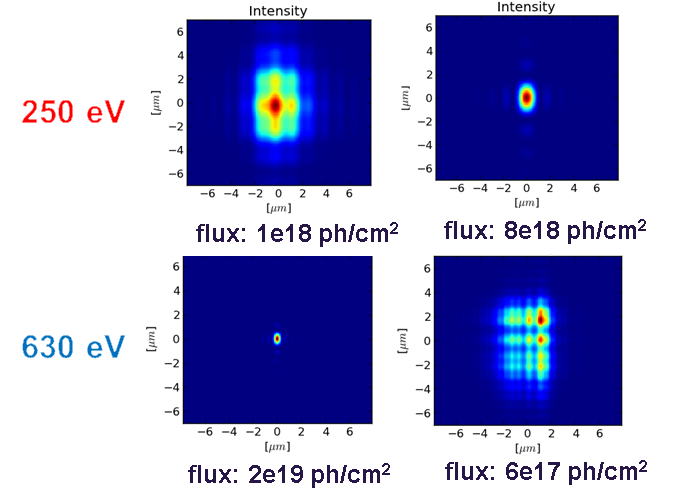}
	\caption{Intensity distributions according to Figure~\ref{fig:2sources} at the locations of the two image planes. Values of the maximum photon flux are provided.}
	\label{fig:2images}
\end{colorfigure}

After passing through the entire beamline, the photon beam can be focused at the sample position. One issue concerning the optimal focusing is related to the presence of the two separate sources for the two pulses, Figure~\ref{fig:2sources}. In our case study the distance between sources $ S1 $ and $ S2 $ amounts to 15~undulator segments, corresponding to a length of about 90\,m. Therefore, only one of the two sources, either at 630 or at 250\,eV can be effectively focused at the sample position (see Figure~\ref{fig:2images}).

In order to ease the issue, one may rely on the flexibility of our setup and open the gap of the last seven undulator sections, instead of the first seven as was shown in Figure~\ref{fig:scheme}. In this case, the distance between the sources $ S1 $ and $ S2 $ is reduced nearly twice - to about 8 undulator segments. This brings down the spatial separation of the images accordingly, but at the cost of a larger clipping of the radiation from the $ S2 $ source.

\begin{colorfigure}[!htb]
	\centering
	\includegraphics[width=0.75\textwidth]{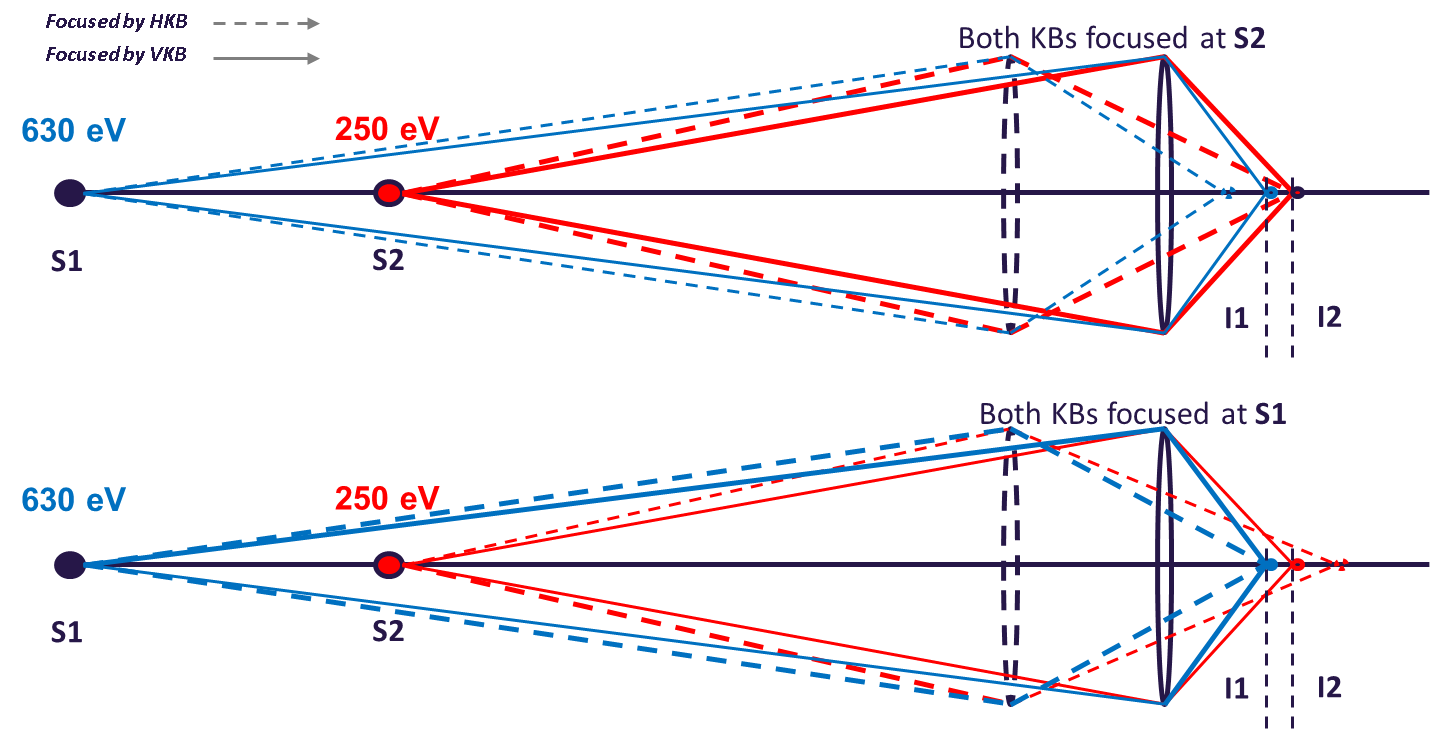}
	\caption{Illustration of astigmatism effects when imaging the two sources: only one source may be reimaged simultaneously in both vertical and horizontal planes.}
	\label{fig:Astigmatism}
\end{colorfigure}
%


It should also be considered that the two KB mirrors have different focal lengths. Therefore, if we choose to obtain an image of $ S2 $ source in both transverse planes (see Figure~\ref{fig:Astigmatism}, top plot), the source $ S1 $ would be imaged at different positions in the horizontal and vertical planes. 
In other words, if we decide to tune the KB mirrors to image one of the two sources at one particular position, the image of the other will not only appear shifted in space, but will also be a subject to astigmatism.

%

It is also possible to select an intermediate, imaginary source position between $ S1 $ and $ S2 $, which we call $ S1.5 $, as in Figure~\ref{fig:compro1}, and image it by properly tuning the KB mirror system. In this case, the image of $ S1.5 $ appears at position denoted $ I1.5 $. This yields a good compromise in terms of beam sizes at the sample, Figure~\ref{fig:compro2}. Based on our simulations, the resulting photon fluxes would be sufficient to conduct the experiment proposed in Section~\ref{sec:sqs_sci_case}.

Since the distance between $ I1 $ and $ I2 $ is about 1\,mm and the molecular beam from the sample delivery can be focused down to 0.1\,mm-order size one may select the location of interaction with high accuracy.

\begin{colorfigure}[!htb]
	\centering
	\includegraphics[width=1\textwidth]{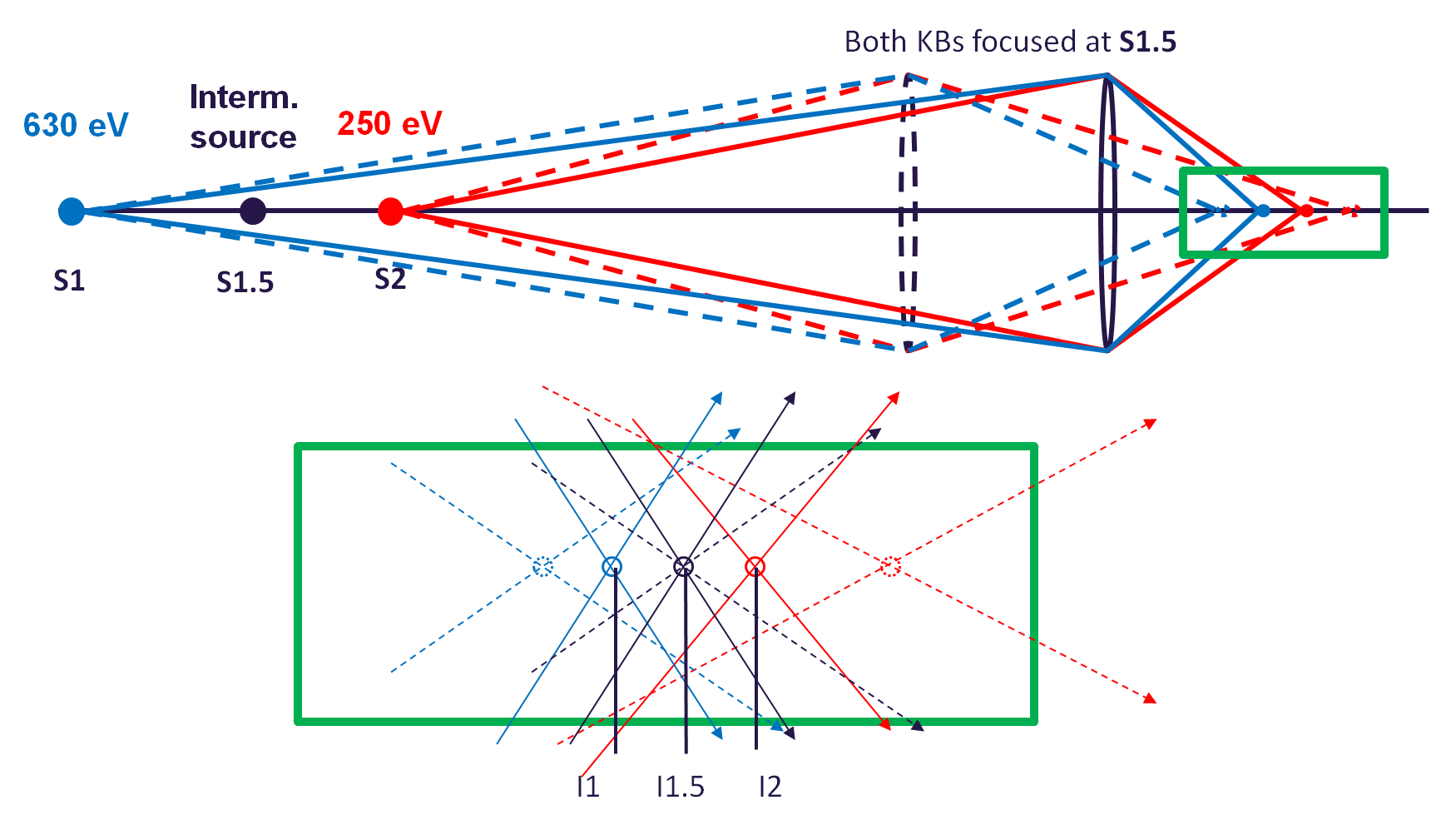}
	\caption{If both KB mirrors are tuned in order to minimize the astigmatism for imaging the virtual source $ S1.5 $, located between sources $ S1 $ and $ S2 $, then the images at planes $ I1 $ and $ I2 $ suffer from aberration. However this solution allows one to obtain a good radiation quality at $ I1.5 $ image plane (see next figure).}
	\label{fig:compro1}
\end{colorfigure}

\begin{colorfigure}[!htb]
	\centering
	\includegraphics[width=1\textwidth]{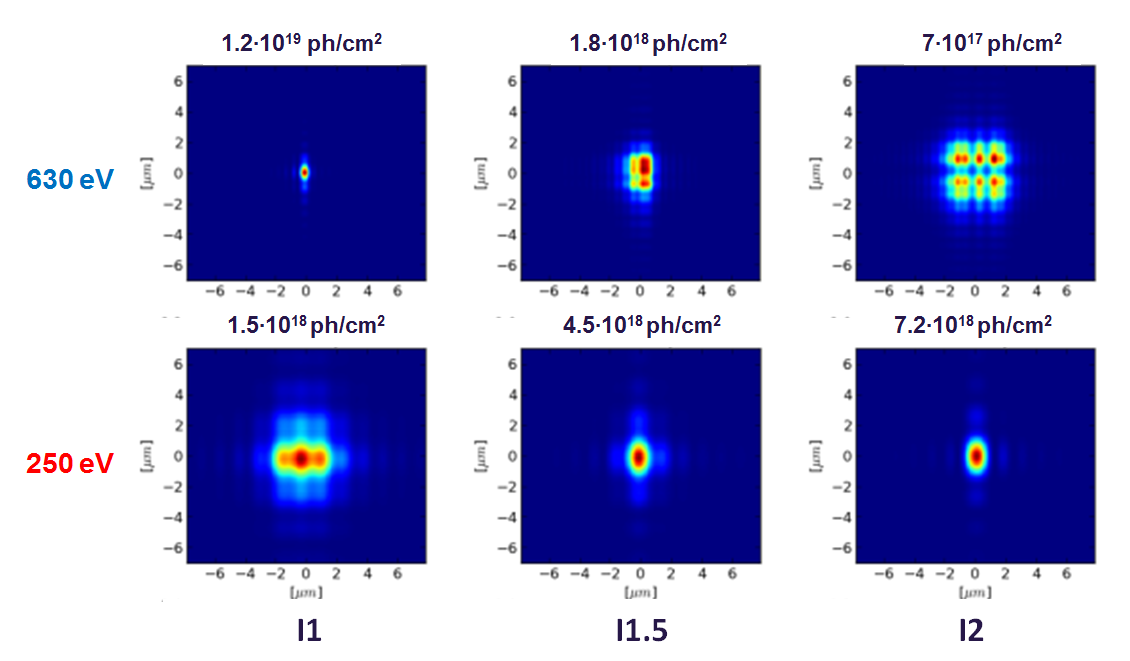}
	\caption{Radiation intensity distribution of both radiation pulses with different photon energies at various image planes from Figure~\ref{fig:compro1}. Peak photon density is provided above the plots. The method of an intermediate source reimaging would allow one to obtain comparable radiation distribution size as well as the photon flux.}
	\label{fig:compro2}
\end{colorfigure}

\section{Conclusions}

There is a great scientific interest to deliver two pulses of different wavelengths with a controllable delay to the experimental samples. In this paper we explore a method to conduct two-color pump-probe experiments at the SASE3 soft X-ray beamline of the European XFEL with minimal hardware modifications.

The scheme, originally proposed at DESY and the European XFEL has been already experimentally tested in LCLS and SACLA and has proven to be robust, easy to set-up and intrinsically resistant to the time delay jitter between the pulses. Both radiation pulses can be satisfactorily focused onto the sample with the baseline KB mirror system. The relative intensity can be varied either by modifying the undulators setup or by changing the settings of the focusing system. Radiation wavelengths can be easily and independently tuned by changing the $ K $ value of the undulators.

This scheme can be technically implemented in two steps: first - installation of an electromagnetic chicane; second - upgrade of the chicane with an introduction of an optical delay line. The second step allows for an increased flexibility of the setup, as it enables to scan through positive and negative delays between the two wavelengths $\lambda_1$ and $\lambda_2$. An option to combine the optical delay with a SXRSS within a single chicane set-up may be also a subject of a future study.

The two-color setup described in this paper can serve both instruments at SASE3, Small Quantum System (SQS) and Spectroscopy \& Coherent Scattering (SCS). In this work we limited ourselves to illustrate our proposal by selecting and analyze one possible application for the SQS instrument, namely the study of charge transfer processes in a linear molecule. We presented the scientific case and we performed start-to-end (s2e) simulations up to the sample position. We started our work from s2e simulations for the electron beam at the entrance of the undulator and we simulated the radiation pulses from our setup. Further on, we computed the propagation of the radiation through the optical beamline up to the focus in the SQS instrument.

While simulations presented in this paper were performed only for the purpose of illustrating the capabilities and the flexibility of the proposed setup, the same computational techniques produce results that may serve as a starting point for detailed simulation of the interaction between radiation and matter, and can be used to define and prepare experiments in great detail.

\section*{Acknowledgments}
We thank Joachim Pfl\"{u}ger, Andreas Scherz and Alexander Yaroslavtsev for useful discussions and Serguei Molodtsov for his interest in this work.
\printbibliography




\end{document}